\documentclass[traditabstract]{aa} 
\usepackage{graphicx,amsmath,amssymb}
\usepackage{txfonts}
\usepackage[squaren]{SIunits} 
\usepackage{multirow}
\usepackage{color}

\newcommand{\au}{\,\hbox{au}}
\newcommand{\microns}{\,\micro\metre}
\usepackage{natbib}
\bibpunct{(}{)}{;}{a}{}{,}

\begin{document}

\title{Collisional modelling of the debris disc around HIP~17439}
 
\author{Ch.~Sch\"uppler\inst{1}
        \and T.~L\"ohne\inst{1} 
        \and A.~V.~Krivov\inst{1} 
        \and S.~Ertel\inst{2}
        \and J.~P. Marshall\inst{3,4}
        \and C.~Eiroa\inst{4}
        }
 
\institute{Astrophysikalisches Institut und Universit\"atssternwarte, 
           Friedrich-Schiller-Universit\"at Jena, 
           Schillerg\"a{\ss}chen~2--3, 07745 Jena, 
           Germany \\
\email{christian.schueppler@uni-jena.de}
 \and
European Southern Observatory, Alonso de Cordova 3107, Vitacura, Casilla 19001, Santiago, Chile 
 \and 
School of Physics, 
University of New South Wales, Sydney, NSW 2052, Australia 
 \and
Departamento de F\'isica Te\'orica, Facultad de Ciencias,
Universidad Aut\'onoma de Madrid,
Cantoblanco, 28049 Madrid, Spain
 }
 
\date{Received \today; accepted ...}
 
  
\abstract 
 {We present an analysis of the debris disc around the nearby K2~V star HIP~17439.  
 In the context of the \emph{Herschel}\thanks{\emph{Herschel} is an ESA space
 observatory with science instruments provided by European-led Principal
 Investigator consortia and with important participation from NASA.} DUNES key programme  
 the disc was observed
 and spatially resolved in the far-IR with the \emph{Herschel} PACS and SPIRE instruments.
 In a first model, Ertel et al. (2014) assumed 
 the size and radial distribution of the circumstellar dust to 
 be independent power laws.
 There, by exploring a very broad range of possible model 
 parameters several scenarios capable of explaining the observations were suggested. 
 In this paper, we perform a follow-up in-depth collisional modelling 
 of these scenarios trying to further distinguish between them.
 In our 
 models we consider collisions, direct radiation pressure, and drag forces, 
 i.e. the actual physical processes operating in debris discs. 
 We find that all scenarios discussed in Ertel et al. are physically sensible and can reproduce
 the observed spectral energy distribution along with the PACS surface brightness profiles
 reasonably well. 
 In one model, the dust is produced beyond $120\au$ in a narrow planetesimal belt and 
 is transported inwards by Poynting-Robertson and stellar wind drag.
 A good agreement with the observed radial profiles would require stellar winds by
 about an order of magnitude stronger than the solar value, which  is not supported~--
 although not ruled out~-- by observations.
 Another model consists of two spatially separated planetesimal belts, a warm inner and
 a cold outer one. 
 This scenario would probably imply the presence of planets clearing the gap between the 
 two components. 
 Finally, we show qualitatively 
 that the observations can be explained by assuming the dust 
 is produced in a single, but broad planetesimal disc with a surface density of solids rising 
 outwards, as expected for an extended disc that experiences a natural inside-out collisional 
 depletion.
 Prospects of discriminating between the competing scenarios by future observations are 
 discussed.  

}
\maketitle


\section{Introduction} 

HIP~17439 (HD~23484) is a nearby solar-type star with an estimated age 
of $0.8$--$3.7\,\text{Gyr}$ \citep{Mamajek2008,Garces2010,Fernandes2011}. 
The first clear evidence for a circumstellar dust disc 
around HIP~17439 was found by \cite{Koerner2010} who detected a far-IR excess 
in \emph{Spitzer}/MIPS data. Recently, the HIP~17439 system was observed 
as part of the \emph{Herschel} \citep{Pilbratt2010} 
Open Time Key Programme DUst around NEarby Stars
\citep[DUNES,][]{Eiroa2010,Eiroa2013}.
The disc appeared spatially resolved in the far-IR images of the 
PACS \citep{Poglitsch2010} and SPIRE \citep{Griffin2010,Swinyard2010} instruments.
Together with the discs around HD~202628 \citep[][]{Krist2012}, 
HD~207129 \citep{Krist2010,Marshall2011,Loehne2012},
and HD~107146 \citep{Ertel2011}, HIP~17439's disc is amongst 
the most extended around sun-like stars identified to date.

Theory predicts that the dust in such a disc
is short-lived compared to the lifetime of the star, owing to
the mutual collisions and radiative forces.
Therefore, the dust is thought to be continually replenished through collisional grinding 
of km-sized asteroidal bodies (planetesimals) that are remnants 
of the planet formation process \citep{Backman1993, Wyatt2008, Krivov2010}. 
In the DUNES programme, those debris discs have been detected with a rate of at least 20\% 
around solar-type 
main-sequence stars \citep{Eiroa2013}.
Their presence may or may not be correlated in one or another way with the presence of 
planets
\citep{Moro-Martin2007, Kospal2009, Bryden2009,%
Maldonado2012, Wyatt2012,Matthews2014,Marshall2014}.

The modelling of the spectral energy distribution (SED) of a debris disc is in 
general a degenerate problem. 
For example, the observed flux can be reproduced by the thermal emission of large dust grains 
close to the star or small ones located far away. 
A spatially resolved debris disc where the location of the dust is directly measurable
offers the best opportunity to break those degeneracies.
Modelling of resolved systems places meaningful constraints on the discs' radial and temperature
structures as well as the grain sizes and compositions 
\citep[e.g.,][]{Matthews2010,Ertel2011,Eiroa2011,Lestrade2012,Booth2013}.

Simultaneous modelling of the SED and the radial surface brightness profiles extracted from the PACS 
images of the HIP~17439 system was done by \cite{Ertel2014} for the first time. 
There, a dust surface number density $n\propto  s^{\gamma} r^{\alpha}$ was assumed, i.e. a combination of 
two independent power laws for the size distribution (exponent $\gamma$) 
and the radial distribution (exponent $\alpha$) of the grains, 
fitted to the observational data by a multi-wavelength $\chi^2$ minimization. 
This way, one can readily explore a huge parameter space and thereby 
find the most appropriate disc configurations.  
\citeauthor{Ertel2014} showed that the SED and the radial profiles 
can be well reproduced by either a one-  
or a two-component scenario (Table~\ref{tab:steves_bestfit}).

\vspace{-0.25cm}

\begin{table}[htb!]
 \centering
 \caption{Best-fit results with $3\sigma$ uncertainties of the power-law models 
          from \cite{Ertel2014}. 
          The dust surface number density is defined as $n\propto s^{\gamma} r^{\alpha} $. 
          $r_1$ and $r_2$ are the inner and outer boundary of a disc, 
          $\theta$ is the disc inclination from face-on,
          and $M_\text{d}$ the dust mass for particles with radii $s<1\milli\metre$.}
 \renewcommand{\arraystretch}{1.5}
 \begin{tabular}{cccc}
 \hline\hline
 Parameter                      & \multicolumn{3}{c}{Best-fit}   \\
 \hline
                                &\multicolumn{1}{c}{One-component model}        & \multicolumn{2}{c}{Two-component model} \\
                                &                                               & Inner disc               & Outer disc \\
 \hline
 $r_1$ [au]                     & \multicolumn{1}{c}{8.3$^{+5.6}_{-0.8}$ }      & 29.2$^{+6.6}_{-27.5}$    & 90.9$^{+79.9}_{-74.9}$\\
 $r_2$ [au]                     & \multicolumn{1}{c}{394.0$^{+106.0}_{-267.4}$} & 500.0$^\ast$             & 500.0$^\ast$ \\
 $\alpha$                       & \multicolumn{1}{c}{-0.1$^{+1.0}_{-1.5}$}      & -4.0$^{+3.6}_{-1.0}$     & -1.6$^{+2.6}_{-3.4}$ \\
 $s_\text{min}$ [$\!\microns$]  & \multicolumn{1}{c}{8.1$^{+2.6}_{-1.9}$}       & 5.2$^{+10.8}_{-1.7}$     & 12.4$^{+17.5}_{-12.3}$ \\
 $\gamma$                       & \multicolumn{1}{c}{-4.0$^{+1.0}_{-0.9}$}      & -5.5$^{+1.8}_{-0.0}$     & -4.3$^{+1.3}_{-1.2}$\\
 $\theta$ [deg]                 & \multicolumn{1}{c}{63.9$^{+18.1}_{-46.1}$}    & \multicolumn{2}{c}{60$^{+10}_{-10}$}\\
 $M_\text{d}$ $[M_\oplus]$      & \multicolumn{1}{c}{$1.3\times10^{-2}$}        & $2.1\times10^{-4}$       & $1.1\times10^{-2}$ \\
 Dust material                  & astrosilicate                                 & \multicolumn{2}{c}{astrosilicate} \\
 \hline   
 \end{tabular}
\label{tab:steves_bestfit}
 \medskip
    \tablefoot{Values marked with $^\ast$ were fixed in the modelling and are not outcomes of the fitting
     procedure.
   }
\end{table}

\vspace{-0.25cm}
 
The best-fit one-component model found is a broad dust disc with a 
radial  extension of several hundreds of au where the number density is nearly constant 
($\alpha=-0.1$). 
Two possible morphologies of such a disc are conceivable. 
The first one is a narrow planetesimal belt near the outer disc edge. Through transport processes 
such as Poynting-Robertson (P-R) drag dust particles move inwards, filling
the inner disc region and ensuring a broad distribution of dust. 
The second possibility is a broad planetesimal ring.
This implies that dust is produced everywhere
in the whole disc, which~-- under certain assumptions~-- may result in a constant dust 
surface density as well. 
The two-component model consists of two rings with inner edges 
around 29 and 91~au. 
Due to the large negative $\alpha$ values, most of the dust is located near the inner disc edges 
of the two components 
and the rings are spatially separated by a wide gap.
Continuing the study by \cite{Ertel2014}, this paper presents an in-depth collisional 
modelling of the HIP~17439 system. 
To this end, we start with an initial distribution of planetesimals, which we also refer to as parent bodies,
and consider their subsequent collisional evolution. As a result, the bodies are ground down to dust 
where the smallest particles end up with typical sizes in the order of $1\microns$ or smaller.
The collisional outcomes depend on the size of the impactors, the material properties, 
and the relative velocities in the disc. 
Our modelling also includes dust transport mechanisms in the form of stellar wind and P-R 
drag, dependent on grain size, material properties and stellar distance.
Hence, the dust radial and size distributions are intrinsically coupled.
Their slopes can no longer be controlled directly, but instead, they are determined by the 
evolution of the planetesimal disc. 
This way, we model the actual physical processes operating in a debris disc. 
However, due to the numerical complexity of this method, 
we can only explore a limited number of parameter combinations. 
As a starting point, we use the one- and two-belt models from \cite{Ertel2014} and 
check whether these are physically plausible.

In Sect.~\ref{sec:model_setup}, we describe the data used and the technique of our modelling. 
Sections~\ref{sec:onering} and \ref{sec:tworings} present simulations with 
one narrow parent belt and two such belts, respectively. 
Section~\ref{sec:extended_belt} discusses the possibility of 
a single extended planetesimal disc. 
Some prospects for future observations are given in Sect.~\ref{sec:prospects}.
Section~\ref{sec:conclusion} contains conclusions and a discussion of our results.


\section{Model setup}
\label{sec:model_setup}
\subsection{Stellar properties}

HIP~17439 is a K2~V star \citep{Torres2006, Gray2006}
at a distance of 16~pc \citep{vanLeeuwen2007}. 
We assume a mass of $0.82~M_\odot$ \citep{Fernandes2011}, 
a bolometric luminosity of $0.40~L_\odot$ \citep{Eiroa2013}, 
and an effective temperature of $5166~\kelvin$ \citep{Eiroa2013}. 
We use a synthetic atmosphere model for the stellar spectrum 
from the PHOENIX/Gaia grid \citep{Brott2005}, 
scaled to WISE and near-IR fluxes \citep[see][]{Ertel2014}.

\subsection{Observational data for the debris disc}

We take the photometric data listed in Table~3 of \cite{Ertel2014}. 
The SED is densely sampled between 12 and $500\microns$ by 
measurements of IRAS, \emph{Spitzer}, and \emph{Herschel}.
The PACS images show an increasing disc extension
from $131\au$ at $70\microns$ to $253\au$ at $160\microns$ along the major axis. 
Throughout this paper, we use the radial profiles derived
by \citeauthor{Ertel2014} from the PACS images at 70, 100, and $160\microns$, but not from the SPIRE images,
since the disc appears only marginally resolved at SPIRE wavelengths and 
the source brightness profile is sparsely sampled, which causes larger 
uncertainties in the radial profiles.

\subsection{Collisional modelling}
We carry out the modelling with our parallelised C++ code \texttt{ACE} that determines the distributions of solids 
in a rotationally symmetric debris disc.
The code follows a statistical approach to solve the kinetic equation
which includes the combined effects of gravity, direct radiation pressure, 
drag forces, and non-elastic collisions 
\citep[for details see][]{Krivov2005, Krivov2006, Loehne2008, Loehne2012}. 
We use the latest version of ACE
that works on a three-dimensional grid with logarithmic bins for object mass, pericentre, and 
eccentricity, as described in \cite{Krivov2013}.

All solids, from dust grains to planetesimals, are assumed to have a spherical shape.
Their sizes are characterised by the radius $s$.
The outcome of each collision depends on the material strength and
relative velocity of the colliders. 
The inclination distribution of bodies in the disc is assumed to be uniform within a constant
semi-opening angle, which is set to half the planetesimal's maximum 
eccentricity $e_\text{max}$ 
\citep[energy equipartition relation, e.g.,][]{Greenberg1991}. 
The eccentricity distribution is also taken to be uniform within
a minimum value close to zero and $e_\text{max}$.
The dust distribution generated by \texttt{ACE} at a certain time instant serves as input for our tools 
\texttt{SEDUCE} and \texttt{SUBITO} that compute the thermal emission and the radial profiles, respectively 
\citep{Mueller2010}. 

We characterise the material strength by the critical specific energy for 
disruption and dispersal $Q_\text{D}^\star$, i.e. the projectile energy per target mass 
necessary to disperse half the target mass:
\begin{align}
 Q_\text{D}^\star = Q_{\text{D,s}} \left(\frac{s}{1\,\metre}\right)^{b_\text{s}} 
\left(\frac{v_\text{imp}}{3\,\kilo\metre \, \second^{-1}}\right)^{0.5}.
 \label{equ:QD}
\end{align}
Equation~(\ref{equ:QD}) follows the formulation of \cite{Benz1999} for 
the strength regime, which we have modified by a power-law dependence on 
the impact speed $v_\text{imp}$ (\citeauthor{Stewart2009} \citeyear{Stewart2009};
see also \citeauthor{Loehne2012} \citeyear{Loehne2012}). 
Because in all our simulations the largest bodies considered in the disc
were $10\,\metre$ in radius, $Q_\text{D}^\star$ is determined solely by the material strength, 
and therefore, we neglect the gravitational binding energy that is usually included 
in the description of 
$Q_\text{D}^\star$. We set $Q_\text{D,s}=10^7 \text{erg} \, \gram^{-1}$ and $b_\text{s}=-0.37$ 
which are close to the values used by different authors 
\citep{Housen&Holsapple1990,Holsapple1994,Benz1999, Stewart2009}.

\subsection{Grain properties}

After collisions the orbits of the newly born fragments are affected by 
radiation pressure which has the net result of increasing the semi-major axes 
and eccentricities of the fragments compared to those of the parent 
bodies. This is determined by $\beta$, the ratio between radiation pressure and 
gravitational attraction. For parent bodies in
nearly circular orbits, the eccentricities of fragments are $>1$ for $\beta>0.5$ 
and they leave the system on unbound orbits 
\citep[e.g.,][]{Burns1979}. 

\begin{figure}[htb!]
 \resizebox{\hsize}{!}{\includegraphics{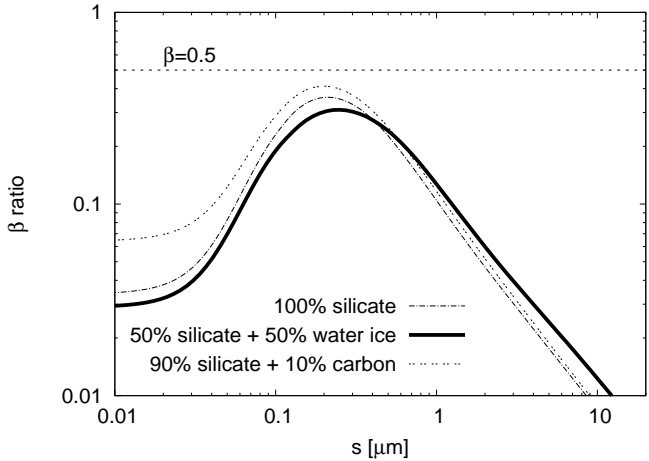}}
 \caption{$\beta$ ratio for several grain materials. 
          The values are calculated with the stellar parameters of HIP~17439.
          Refractive indices are taken from \cite{Draine2003b}
          (astrosilicate), \cite{Li1998} (water ice), and \cite{Zubko1996} (amorphous
          carbon).}
 \label{fig:beta}
\end{figure}

Figure~\ref{fig:beta} compares $\beta$ as a function of the dust grain 
radius $s$ for different materials. For all compositions assumed here, 
$\beta = 0.5$ is not reached in the HIP~17439 system and the 
blowout limit does not exist.
For the modelling in this paper, we choose
the 50:50 mixture of astrosilicate 
and water ice with a density of $2.35\,\gram\, \centi\metre^{-3}$. 
The optical constants of this mixture are determined via effective medium theory 
and the Bruggeman mixing rule \citep{Ossenkopf1991}. 
The absorption and radiation pressure efficiencies
are calculated by means of Mie theory \citep{Bohren1983}.
These techniques have been used for the modelling 
of other discs in the \emph{Herschel}/DUNES \citep[e.g.,][]{Loehne2012},
GASPS\footnote{ GAS in Protoplanetary Systems \citep{Dent2013}} \citep[e.g.,][]{Lebreton2012}, 
and DEBRIS\footnote{Disc Emission via Bias-free Reconnaissance in the Infrared/Submillimetre 
(\citeauthor{Phillips2010} \citeyear{Phillips2010}; Matthews et al. in prep.)} \citep[e.g.,][]{Churcher2011}
programmes. 

\section{One parent belt}
\label{sec:onering}

At first we analyse the scenario where dust is produced in a single narrow planetesimal ring 
far away from the host star. Due to transport processes through P-R drag and stellar wind drag, 
the dust grains move inwards and fill the inner disc region.
A similar scenario was considered for $\varepsilon$~Eri by \cite{Reidemeister2011} 
who showed that warm dust close to the star could stem from an outer 
Kuiper belt analogue between 55 and $90\au$.

\subsection{Model description}
\label{sec:model_descr}

We performed several \texttt{ACE} runs with input parameters listed in 
Table~\ref{tab:input_onering}. 
The parameter space was sampled by 35 bins in mass from 
$5\times 10^{-18}\gram$ to $1\times10^{10}\gram$  
(i.e., $0.01\microns$ to $10\,\metre$ in size)
and 21 bins in pericentric distance 
from 9 to $250\au$. The eccentricity grid covered values 
from 0.001 to 2.5 with 25 bins.

After a certain timestep the simulations reach an equilibrium state in which 
the amounts of particles with different sizes on different orbits stay constant 
relative to each other. Once the equilibrium state is reached, the 
dust mass in the systems decreases continually with time. We stopped the simulations
as soon as the dust's thermal emission has reduced to the observed excess strength, 
given by the photometric data. 
Because the simulation times $t_\text{sim}$ depend strongly on the choice of the 
initial conditions, $t_\text{sim}$ do not necessarily correspond to the system's 
physical age $t_\text{phys}$.
Hence, $t_\text{sim}$ can possibly exceed $t_\text{phys}$. For instance, if we 
start with a massive planetesimal belt, much dust is produced and it takes  
time to reach a rather low excess level. 
For all runs in Table~\ref{tab:input_onering}, the initial planetesimal populations were 
uniformly distributed between $r_{\text{p},1}$ and $r_{\text{p},2}$ with 
size distribution index $\gamma=-3.5$ and initial masses of $\approx\!1~M_\oplus$. 
The corresponding $t_\text{sim}$ turned out to be $\lesssim\!2$~Gyr, longer 
than the relaxation times ($\lesssim\!500$~Myr) needed to the reach the 
equilibrium state.

\begin{table}[h]
 \centering
 \caption{\texttt{ACE} input parameters for one-belt models.}
 \label{tab:input_onering}
 \begin{tabular}{lccccc}
  \hline\hline
    Model   & Drag                       & $r_{\text{p},1}$   & $r_{\text{p},2}$     & $e_\text{max}$ & $\theta$  \\
            &                            & [au]               & [au]                 &                & [deg] \\
  \hline
  FG        & P-R only                   & 120                & 150                  & 0.04           & 65 \\
  SW1       & P-R\textbf{+15 SW$_\odot$} & 120                & 150                  & 0.04           & 65 \\
  SW2       & P-R\textbf{+30 SW$_\odot$} & 120                & 150                  & 0.04           & 65 \\
  OSP       & P-R\textbf{+15 SW$_\odot$} & \textbf{150}       & \textbf{180}         & 0.04           & 65 \\
  \hline
 \end{tabular}
 \medskip
     \tablefoot{
          SW$_\odot$ denotes stellar winds of solar strength, 
          $e_\text{max}$ the maximum eccentricity of the initial distribution of planetesimals, and
          $\theta$ the disc inclination from face-on.  
          $r_{\text{p},1}$ and $r_{\text{p},2}$
          are the inner and outer edge of the planetesimal belt.          
          Changes with respect to the first-guess (FG) model are marked in bold.
          For further description of the models see text.      
   }
\end{table}

\subsection{First-guess modelling}
\label{sec:FG_model}

\begin{figure}[htb!]
 \resizebox{\hsize}{!}{\includegraphics{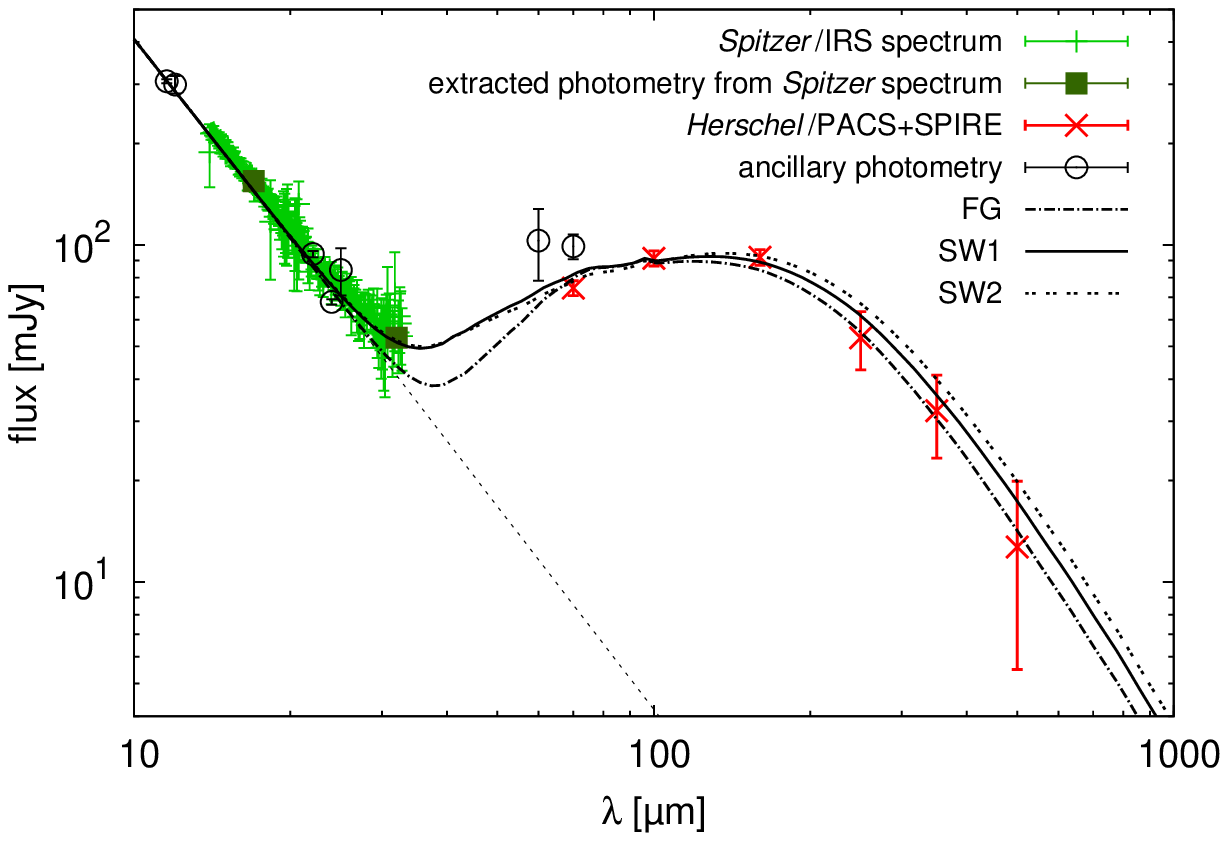}}
 \resizebox{\hsize}{!}{\includegraphics{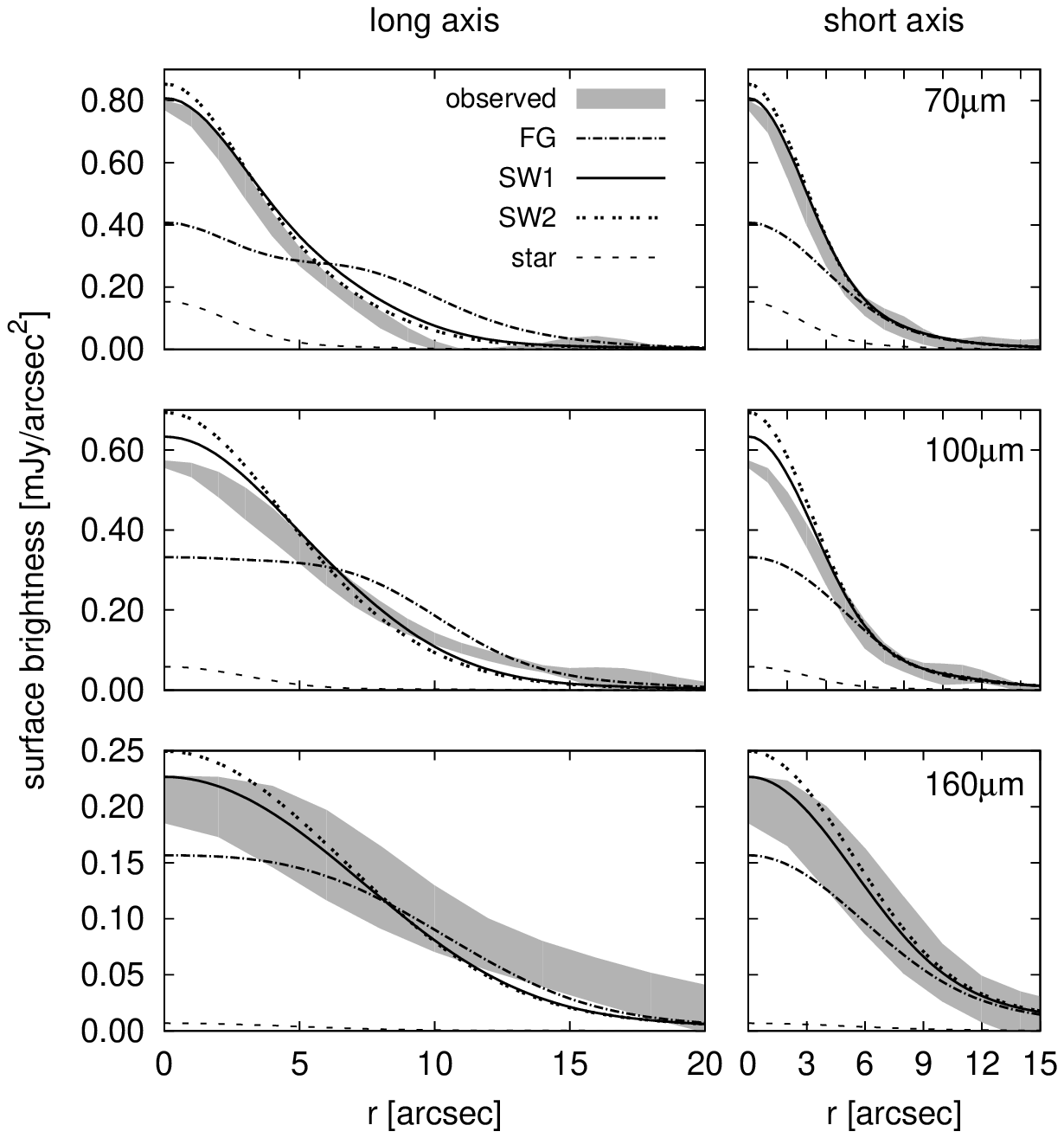}}
 \caption{Observed SED and radial profiles plotted along with the simulated data  
          from the FG, SW1, and SW2 models in Table~\ref{tab:input_onering}. 
          For a detailed description of the observational data see \cite{Ertel2014}.
          Note that data points with a signal-to-noise ratio $<3$           
          are excluded from the available \emph{Spitzer}/IRS spectrum.          
          In the radial profile plots, the gray-shaded areas depict the observed
          data with $1\sigma$ uncertainties at 
          PACS wavelengths (from top to bottom: 70, 100, and 160$\microns$). 
          Cuts along the long and short axes
          are given in the left and right column, respectively.  
         }
 \label{fig:onering_SED+radprof}
\end{figure}

We started with a first-guess (FG) run where we assumed 
a location of the planetesimal belt between $120$ and $150\au$, 
which is at the lower boundary of the 3$\sigma$ confidence level for the outer radius of 
the one-component  model of \cite{Ertel2014} (Table~\ref{tab:steves_bestfit}). 
The dust particles are transported by P-R drag from their 
birth places towards the star. 
Figure~\ref{fig:onering_SED+radprof} shows the resulting SED and radial profiles.

The SED reproduces the observations quite well, except that 
the mid-IR flux probed by \emph{Spitzer}/IRS is underestimated. 
Although at $160\microns$ the radial profiles are almost within the 
$1\sigma$ uncertainties of the observational data, they are distinctly
too shallow for $r<7{''}$ at 70 and $100\microns$.
This clearly illustrates a deficit of particles 
in the inner part of our modelled disc.  We expect that a conceivable relocation of 
the initial planetesimal belt would not lead to significant improvements. 
If the belt is shifted inwards, the averaged collisional velocities will increase 
($v_\text{imp}\propto r^{-0.5}$), thus lowering the collisional lifetime of larger grains. 
The amount of larger particles would be reduced, and as a result, the
submillimetre emission is depressed.  
That would have a detrimental effect because the FG model is already well fitted to the 
submillimetre emission. Conversely, shifting the belt outwards would hamper 
the collisional depletion of larger particles and therefore the production of smaller ones. 
Hence, even fewer small particles would be transported into the inner region. 
On the basis of a narrow parent belt, an 
increase of the inner disc emission without changing the 
submillimetre flux substantially can only be possible 
by strengthening the inward transport of dust and not by shifting the belt.

\subsection{Adding stellar winds} 
\label{sec:adding_winds}

Stellar winds cause dust grains to lose angular momentum due to 
collisions with the wind particles, thereby enhancing the P-R drag.
Accordingly, the P-R timescale for the inward transport 
is shortened by a factor of $\left[1+(\beta_\text{SW}/\beta)\, 
(c/v_\text{SW})\right]^{-1}$ \citep{Mukai1982}, 
where $c$ and $v_\text{SW}$ are the speed of light and 
of the stellar wind particles, respectively. 
Here, $\beta_\text{SW}$ denotes the ratio between stellar wind pressure and gravity, 
which is proportional to the stellar mass loss rate $\dot{M}$ \citep{Minato2006}.

Many cool, late-type stars on the main sequence are known to possess
strong stellar winds \citep{Wood2002,Wood2005}.
For moderately rotating K main sequence stars with radius $R_\star$,
\cite{Wood2005} found a correlation between 
the stellar X-ray flux $F_\text{X}$ and the mass loss rate by $\dot{M_\star}/R_\star^2 \propto F_\text{X}^{1.34\pm0.18}$ 
as long as $F_\text{X} \lesssim 8\times 10^5\, \text{erg} \, \second^{-1} \centi\metre^{-2}$.
For HIP~17439's X-ray luminosity of $\log L_\text{X} /L_\star =-4.9$ \citep[ROSAT all-sky survey bright source catalogue,][]{Voges1999}
and $R_\star=0.8\,R_\odot$,  
this yields $24.6^{+15.6}_{-9.5} \dot{M}_\odot$.
On the other hand, 
the theoretical prediction of $\dot{M}_\star$ by \cite{Cranmer2011} for HIP~17439's mean rotation period
of 11 days is significantly smaller, about $5\,\dot{M}_\odot$  
\citep[see Fig.~13 in][]{Cranmer2011}.
Furthermore, there are no clear observational indicators that HIP~17439 
is active enough to have strong stellar winds (J. Sanz-Forcada, private 
communication).  
Therefore, a strong stellar wind assumption should be seen as critical. 
Nevertheless, we ran simulations with stellar winds to probe their effects 
on the dust distribution and to test whether the model can be improved this way.

We started two stellar wind runs, SW1 and SW2, where 
we assumed $15$ and $30~\dot{M}_\odot$, respectively.
The SW1 mass loss rate corresponds to the lower limit of 
the \citeauthor{Wood2002} criterion mentioned above. The SW2 one
is equal to what has been proposed for $\varepsilon$~Eri \citep{Wood2002}, 
a star that is similar to HIP~17439 in mass, 
radius, and X-ray luminosity \citep{Benedict2006, DiFolco2007,Schmitt&Fleming1995}. 
Although there is still a lack of observations, 
cool K-dwarfs are expected to have hot coronae and stellar winds that are driven 
by gas pressure similar to solar winds \citep[e.g.,][]{Lamers1999,Wood2004}. 
Therefore, we assume $v_\text{SW}=400~\kilo\metre \, \second^{-1}$  
equal to the averaged solar wind speed in SW1 and SW2.

\begin{figure*}[htb!]
 \sidecaption
 \includegraphics[width=12cm]{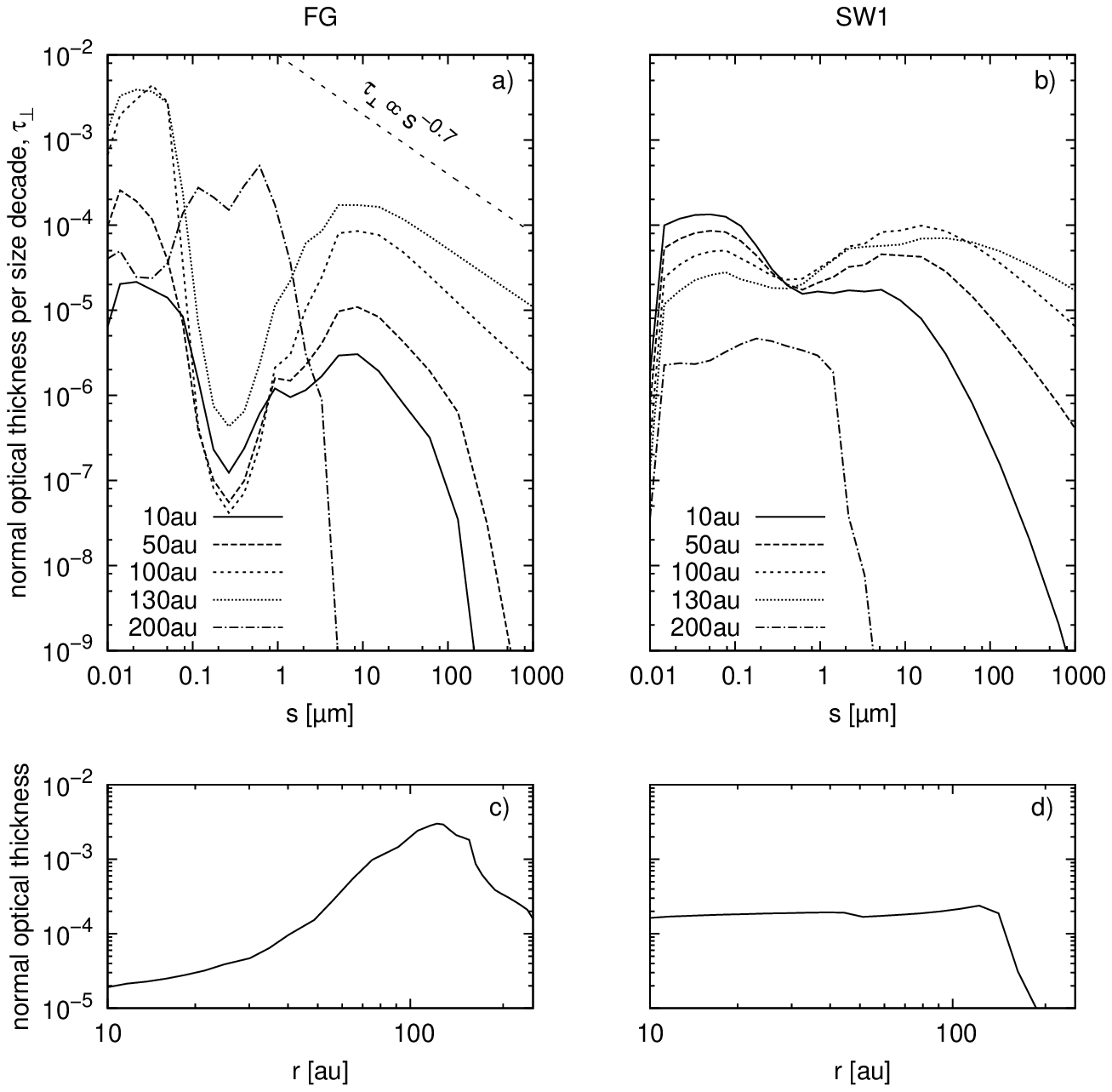}
 \caption{Size and radial distributions of particles in disc models
          FG (\emph{left column}) and SW1 (\emph{right column}).                                                              
          \emph{Top:} Normal optical thickness per size decade $\tau_\perp$
          as a function of grain size $s$.
          The curves depict the contributions at several distances from the star.
          For comparison, the straight line in panel a) 
          shows the theoretically
          predicted slope for an infinite collisional cascade assuming
          $Q_\text{D}^\star \propto s^{b_\text{s}}$ with $b_\text{s}=-0.37$,
          which is $n(s)\propto s^{-3.7}$ \citep[][]{OBrien2003}. 
          \emph{Bottom:} Radial profiles of the 
          total optical thickness (integrated over all grain sizes).
          }
 \label{fig:onering_tau}
\end{figure*}

Figure~\ref{fig:onering_SED+radprof} demonstrates that the SEDs of the stellar wind 
runs predict more emission between 20 and $70\microns$ compared to FG. 
The radial profiles become steeper with increasing stellar wind strength
and considerably better match the observational data.
This is especially the case for SW1, whereas the bigger deviations from the data 
in model SW2 show that the winds are already too strong there.
The reasons for the SED and radial profile modifications through stellar winds
can be better understood by comparing the size distributions of the different models. 
For this purpose, Figs.~\ref{fig:onering_tau}a,b show the normal
optical thickness per size decade $\tau_\perp (s) \propto n(s) \cdot s^3$ for FG and SW1.

\begin{figure*}[htb!]
 \sidecaption
 \includegraphics[width=12cm]{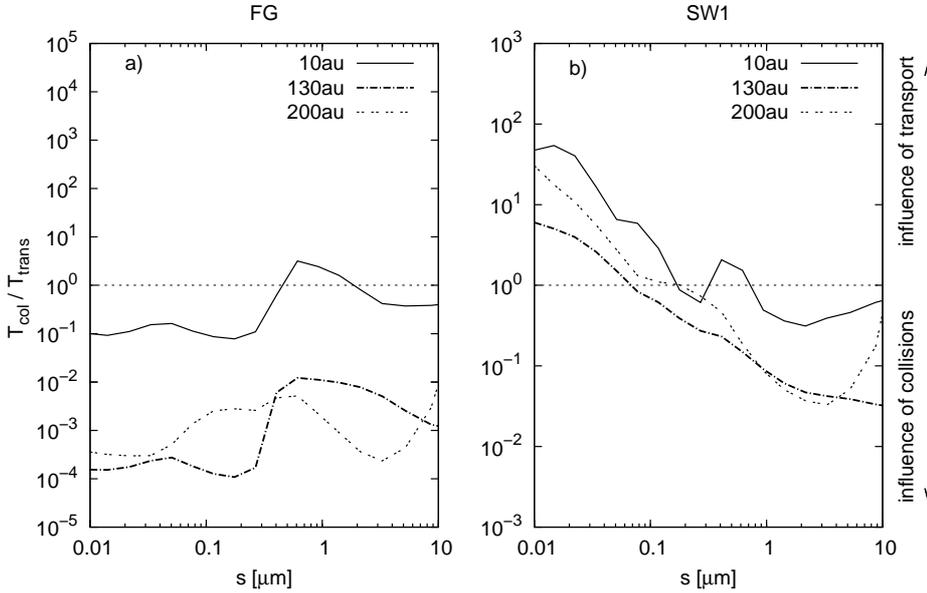}
 \caption{Ratio of the collision and transport timescales of the FG 
          and SW1 run inside ($10\au$), within ($130\au$), 
          and outside ($200\au$) the planetesimal ring. 
          Only particles with eccentricities $e<0.02$ are included.  
          }
 \label{fig:onering_timescales}
\end{figure*}

The amount of particles with $s>1\microns$ is clearly 
increased inside the planetesimal belt in run SW1 (curves for $r<130\au$ in Fig.~\ref{fig:onering_tau}b 
are more extended towards bigger sizes compared to those of the FG model). 
Furthermore, the presence of stellar winds causes $\tau_\perp$ to go sharply down for $s<0.02\microns$ 
because $\beta + \beta_\text{SW} > 0.5$ and 
the grains are blown away.  
In the FG model there is a deep minimum of $\tau_\perp$ around $s=0.3\microns$ within
and interior to the parent belt (see Fig.~\ref{fig:onering_tau}a, lines for $r\leq130$~au). Since $\beta$ is highest around $s=0.3\microns$ 
(Fig.~\ref{fig:beta}), these particles are pushed on eccentric orbits 
after their collisional production
and contribute little to the dust density 
in the inner disc region. In the SW1 model this minimum is markedly shallower 
because small particles 
move quickly towards the star and thereby 
are less efficiently destroyed by collisions.

To confirm the last statement, we also checked the role of transport 
by evaluating the collisional and transport lifetimes  
(Fig.~\ref{fig:onering_timescales}). 
We define the collisional timescale as 
$T_\text{col} = N_i / \dot{N}_i$. Here, $N_i$ denotes the number of particles per mass, 
pericentre, and eccentricity bin $i$ (where $i$ is a three-dimensional multi-index).  
$\dot{N}_i$ is the particle loss rate due to disruptive and cratering collisions in 
bin $i$, determined by ACE.\footnote{$\dot{N}_i$ describes the middle term 
with the loss function $L_{ij}$ of Eq.~(8) in \cite{Loehne2012}.} 
We calculated $T_\text{col}$ only for particles in the lowest eccentricity bin whose 
pericentres are roughly equal to their mean orbital distances. 
Using the change rate of orbital elements under the 
influence of a drag force given in \cite{Wyatt1950}, we estimate 
the transport time for particles on
low-eccentricity orbits as
\begin{align}
 T_\text{trans}= 400 \left(\frac{r}{\au}\right)^2 \left(\frac{M_\odot}{M_\star}\right) 
                 \frac{1}{\beta} \left[ 1+ 750 \, 
                 \left( \frac{\beta_\text{SW}}{\beta}\right)  \right]^{-1} \text{yr} . 
 \label{equ:T_trans}
\end{align}
Equation~(\ref{equ:T_trans}) describes the time it takes for a particle to 
spiral inwards from a certain distance $r$ onto the star ($r=0$). 
If $T_\text{col}\gg T_\text{trans}$, the particles are most affected by transport, 
for $T_\text{col} \ll T_\text{trans}$ by collisions.

Comparing Figs.~\ref{fig:onering_timescales}a and \ref{fig:onering_timescales}b 
shows now that small particles in the SW1 run are more affected by transport than collisions. 
Accordingly, the number of collisionally produced fragments on eccentric orbits reduces 
and $\tau_\perp$ tends to flatten from $s=10\microns$ down to the smallest 
grains. The inner region of the planetesimal ring is more strongly
filled by dust and has a nearly uniform density profile proportional to $r^0$ 
(Fig.~\ref{fig:onering_tau}d).
Consequently, the excess emission for $\lambda<70\microns$ increases
and the radial profiles steepen.

\subsection{Shifting the planetesimal belt}

\begin{figure}[htb!]
 \resizebox{\hsize}{!}{\includegraphics{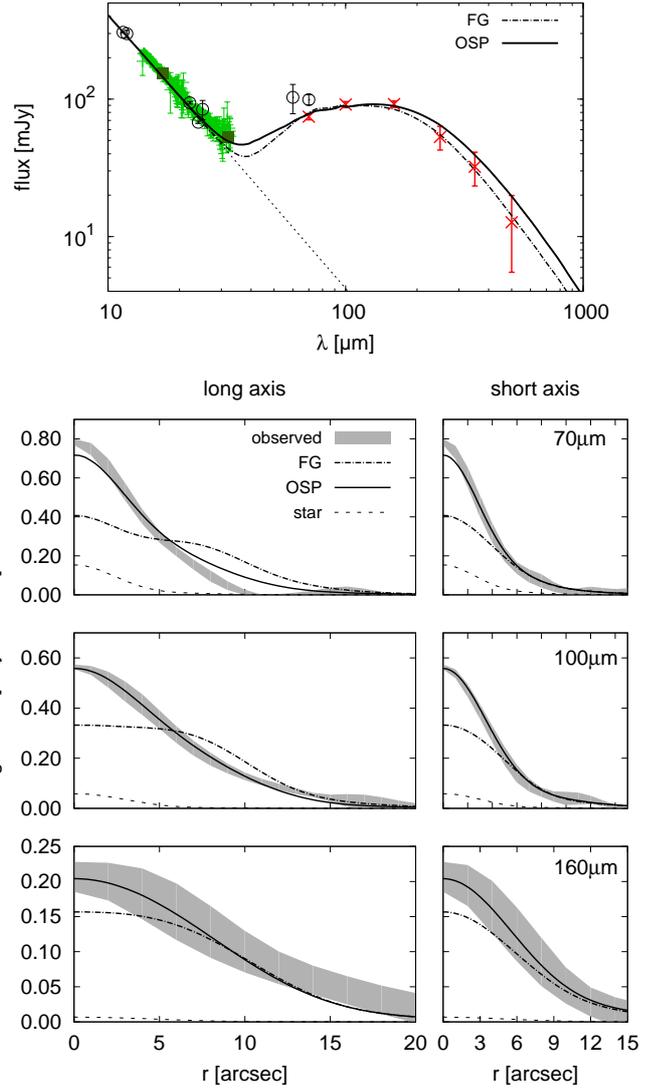}}
 \caption{Same as Fig.~\ref{fig:onering_SED+radprof}, but for FG and OSP (Table~\ref{tab:input_onering}).}
 \label{fig:onering_15SW_d+}
\end{figure}

The previous section illustrated that adding stellar winds with a moderate 
strength of 15 times the solar value can significantly 
improve the model. However, the long axis radial profiles at 100/160$\microns$
are below the observed emission outside the planetesimal ring ($r>10{''}$),
whereas they tend to overestimate it for $r<5{''}$. There are at least 
two modelling options for reducing the surface flux close to the star and simultaneously 
enhancing it farther out:
1. lowering the stellar wind strength (compare SW1 and SW2 in Sect.~\ref{sec:adding_winds}) or 
2. shifting the planetesimal ring outwards (see discussion in Sect.~\ref{sec:FG_model}). 
We explicitly tested the second possibility in a model with outwards shifted planetesimals (OSP).
As expected, the 100/160$\microns$ profiles flatten and better match the observational 
data compared to those of the SW1 model (Fig.~\ref{fig:onering_15SW_d+}). 
We found a fractional luminosity of $9.5\times10^{-5}$ and a dust mass 
(particles with radii below $1\milli\metre$) of $9.5\times10^{-3}M_\oplus$ for the OSP run.

At this point, we refrain from further refinements of the one-disc model.
It would certainly be possible to find a yet better fit to the data
than the one shown, but due to the time-consuming collisional 
modelling (e.g., run OSP took about 72~CPU days) we have to restrict our analysis to a few test runs.
We deem the results presented here
sufficient to illustrate the feasibility of the one-belt scenario.

\section{Two parent belts}
\label{sec:tworings}

As shown by \cite{Ertel2014}, the one-component model in Table~\ref{tab:steves_bestfit} fits the data in general 
with reasonably low $\chi^2$, but the $100\microns$ major axis profile is not sufficiently extended,
leading to the biggest deviations between the observational data and this model
\citep[see Fig.~3 in][]{Ertel2014}.
All modifications that cause a broader emission at $100\microns$ would also result in a substantially broader emission 
at $70\microns$ that is not observed.
This problem motivated a two-component model where the 
emissions at $70$ and $100\microns$ can be adjusted independently if the inner disc mostly contributes to 
$70\microns$ but marginally to $100\microns$, and vice versa 
for the outer disc.

We now attempt to find an appropriate two-component disc by collisional modelling. 
Similar to the results of \citeauthor{Ertel2014}, 
we are searching for an outer disc that dominates 
the flux at $160\microns$, but simultaneously has less influence at shorter wavelengths. 
Consequently, the typical grain size of the dust have to be increased,
otherwise the thermal emission would be too warm. 
This can be achieved if the dynamical excitation of the planetesimals in a disc is low.
Due to the low collisional velocities between the largest bodies, the production 
of small dust grains is inhibited.
On the other hand, the destruction rate of small particles 
is high because they have large eccentricities induced by radiation pressure. 
This imbalance would clean out the disc of small particles \citep{Thebault2008}.

\subsection{Model description}
We placed two narrow, well-separated planetesimal belts and modelled the evolutions of 
the produced dust independently. 
To this end, we started \texttt{ACE} twice -- once only for the inner and once only for the outer belt -- 
and combined the final simulation results. 
We manually determined the time steps $t_\text{sim}$ in the evolution of inner 
and outer component where the sums of their SEDs and radial profiles match 
the observational data the best. 
Because the evolution of both discs are not coupled in any way,
we chose $t_\text{sim}$ independently for inner and outer component.

We tested two different combinations of an inner and an outer disc 
(model~I and model~II), listed in Table~\ref{tab:input_two}. In all models 
the particles were transported by P-R drag only. We used the same mass and eccentricity 
resolution as described in Sect.~\ref{sec:model_descr}. The pericentre grid was set 
from 1 to $100\au$ for the inner discs and from 10 to $200\au$ for the outer discs, 
each with 21 bins. Under the same initial conditions given in Sect.~\ref{sec:model_descr} we found  
$t_\text{sim}\lesssim1$~Gyr and $t_\text{sim}\lesssim8$~Gyr 
for the inner and outer discs, respectively. 
Note again the simulation times merely reflect the initial conditions
of the \texttt{ACE} models. If they are chosen inappropriately, $t_\text{sim}$
can exceed the physical age of the HIP~17439 system 
(see Sect.~\ref{sec:model_descr}).

\begin{table}[h]
 \caption{\texttt{ACE} input parameters for two-belt models. For parameter declarations see Table~\ref{tab:input_onering}.}
 \centering
 \begin{tabular}{ccccccccc}
  \hline\hline
  Model                  &  Disc 	& $r_{\text{p},1}$ 	& $r_{\text{p},2}$  &  $e_\text{max}$ & $\theta$  \\
                         &	        &[au]            	& [au]              &                 & [deg]     \\
  \hline    
  \multirow{2}{*}{I}     &  inner	& 30			& 40	            & 0.04            & 65        \\
                         &  outer       &100			&130		    & 0.001           & 65        \\
  \hline 
  \multirow{2}{*}{II}    &  inner	& 30			& 40	            & 0.04            & 65        \\
                         & outer        &150			&180		    & 0.04            & 65        \\ 
  \hline
 \end{tabular}

   \label{tab:input_two}
\end{table}

\subsection{Results}
\begin{figure}[htb!]
 \resizebox{\hsize}{!}{\includegraphics{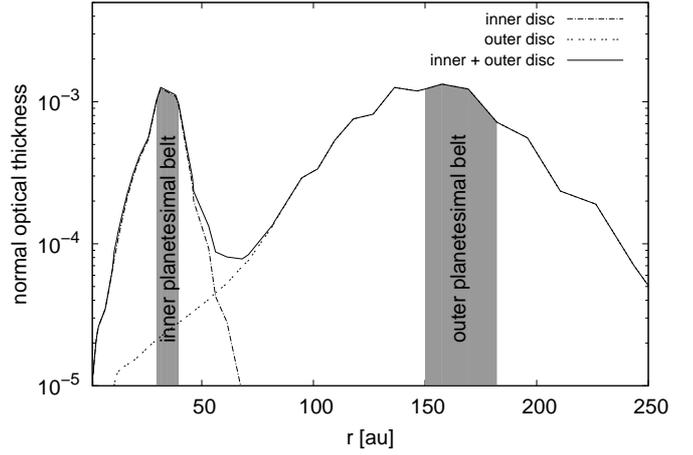}}
 \caption{Radial profile of the optical thickness in model II. 
          The shaded regions mark the extent of the inner and outer parent belts. }
 \label{fig:tau_tworings}
 \end{figure}

Although the parent rings of the inner and outer disc are separated by $\gtrsim 100\au$,
there is certain overlap of the local dust densities, both due to particles transported inward
from the outer disc and the halo grains of the inner one. 
However, judging by the simulations, in model~I and II the outer disc optical thickness
is more than one order of magnitude lower at the location of 
the inner planetesimal belt, and the inner disc optical thickness is 
several orders of magnitude lower at the position of the
outer planetesimal belt (Fig.~\ref{fig:tau_tworings}).
Thus, inner and outer discs can be seen as collisionally decoupled and their independent treatment 
is well justified. 

\begin{figure*}[htb!]
\centering
\includegraphics[width=17cm]{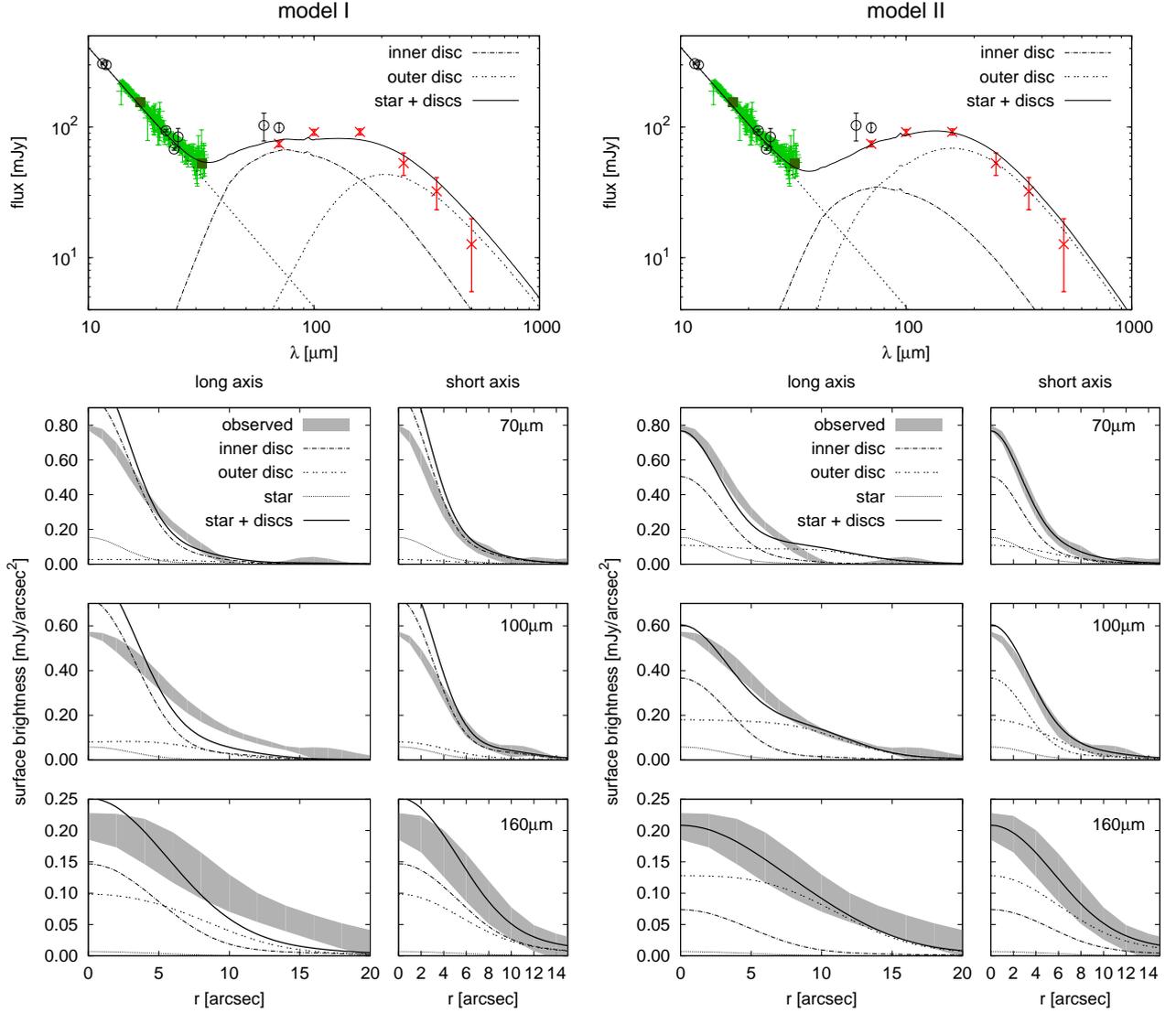}
\caption{SEDs and radial profiles for model I \emph{(left column)} 
         and model II \emph{(right column)} in Table~\ref{tab:input_two}.  
         Profiles for inner and outer discs are star-subtracted.}
\label{fig:two-comp}
\end{figure*}

In model~I,  we set the planetesimal belts close to the locations of the 
inner edges of \citeauthor{Ertel2014}'s two-component model. 
Our first guess for $e_\text{max}$ of the outer disc was 
0.001, quite a low value, inspired by the modelling in \cite{Krivov2013} for 
the \emph{Herschel} cold disc candidates, and distinctly smaller than $e_\text{max}$ 
for the inner disc ($0.04$). 
The results are presented in Fig.~\ref{fig:two-comp}.
The SEDs of both components are rather separated, i.e. the outer disc SED peaks 
beyond $160\microns$ and has minor contribution to 70 and $100\microns$ compared 
to the inner one.  
The overall SED clearly shows emission that is too low at 100 and $160\microns$,
whereas it tends to overestimate the SPIRE photometry by $\approx\!1\sigma$.
Putting more weight on the outer disc SED, i.e. taking the dust mass from an 
earlier time step, 
consequently  improves the agreement with the 100 and $160\microns$ points but has 
the opposite effect at longer wavelengths. The radial profiles 
have a steep fall-off that contradicts the observations. 
This is because they are mostly affected by the inner component, which is only
marginally resolved.

A significant improvement of the model requires the radial profiles
at 100 and $160\microns$ to be broadened without simultaneously increasing the flux at SPIRE wavelengths,
equivalent to a higher contribution of the outer, colder component.
To this end, we departed from the \citeauthor{Ertel2014} setup, as it is difficult to
translate their findings directly to a consistent collisional model.
Both components of their best-fit model have large minimum grain sizes
and size distributions with exponents $\gamma<-4$ (Table~\ref{tab:steves_bestfit}). 
These steep size distributions cannot be explained by modelling with standard collisional prescriptions.
The resulting size distributions from statistical codes such as \texttt{ACE} are typically broader,
leading to broader SEDs as well. The entanglement in terms of resolvable emission from the two
components in our model~I is thus greater than in \citet{Ertel2014}, reducing the ability
to adjust the radial profiles at $70\microns$ and $100/160\microns$ independently.

In our model~II, we therefore shifted the outer planetesimal belt farther away from the 
star and additionally increased $e_\text{max}$ up to 0.04 for this component.
Compared to model~I, the outer disc SED peaks at about $160\microns$ and is markedly
more extended towards shorter wavelengths (Fig.~\ref{fig:two-comp}).
Hence, the proportion changes, i.e. the outer disc is more weighted 
and largely determines the overall SED at $\lambda > 100\microns$. 
Model~II fits the PACS photometric points well, but still shows the trend of 
overestimating the SPIRE photometry.
Thanks to the stronger contribution of the outer disc, all profiles are sufficiently 
broadened. For this inner/outer discs combination we found  
dust masses of $M_\text{d,inner}=4.4\times10^{-4}M_\oplus$ and 
$M_\text{d,outer}=8.7\times10^{-3}M_\oplus$.

Here again, we stress that from numerous parameters, 
such as the position and extent of the parent 
belts as well as their maximum orbital eccentricities, 
we explicitly tested a few ones only. 
The already good fitting result of model~II, which
can surely be improved by further parameter variations,
demonstrates that a two-belt model is capable of explaining 
the data.


\section{One extended parent belt}
\label{sec:extended_belt}

Alternatively to the one narrow belt and the two-belt model, \cite{Ertel2014} also 
suggested that the data may be compatible with one extended parent belt. Such a disc 
with planetesimals distributed over a range from a few au to a few hundreds of au cannot 
be treated with \texttt{ACE} because it would need a prohibitively long computation time. 
However, the results from our two ring model give us a clue for the structure of an 
extended belt model. If we added a third ring between the two existing ones, the resulting 
system would serve as a proxy for an extended planetesimal belt configuration. 
Surely, an intermediate ring would lead to the effect that the belts are not longer 
collisionally decoupled. \cite{Thebault2007} considered a comparable situation 
where they divided an extended debris disc into several annuli and studied their collisional
evolution. Inter-annuli interactions were taken into account by the transport of small 
grains due to radiation pressure. Therefore, particles within a certain annulus can 
collide with high-$\beta$ grains produced in another annulus closer in. Because the impact 
velocities in such events are higher compared to the ones between locally produced objects 
of the same size, the number of destructive collisions increases. We expect the same 
effect in ACE simulations for a setup of adjacent planetesimal rings in the HIP~17439 system.
Thus, the amount of particles on eccentric orbits, which have high $\beta$'s but always
$<0.5$, increases compared to the case of non-interacting rings. Consequently, the overall 
size distribution in the inner disc regions would have a more pronounced minimum at 
$s\!\approx0.3\microns$ where $\beta$ is greatest for our chosen dust material. Thus, a 
simplified assumption of several non-collisionally interacting rings, instead of a real 
extended disc, will lead to some shortcomings that mainly become apparent in the
excess strength at short wavelengths, but are rather unimportant for a qualitative 
discussion of this model.

Figure~\ref{fig:tau_tworings} shows that the peak values of the optical thickness 
of inner and outer ring are roughly the same, about  $1.4\times10^{-3}$. 
A conceivable additional planetesimal ring between the two existing 
would increase the dust density at $r\approx\!70\au$, flattening the overall optical 
thickness. For such a ``quasi-extended'' planetesimal ring we predict a size distribution 
index of $\approx\!-3.7$ as it was found in Sect.~\ref{sec:onering} within the narrow 
parent belt of the FG model. This would be close to the value (-4.0) found by 
\cite{Ertel2014} for the extended one-component model.

An extended planetesimal belt scenario could naturally be expected, for instance, if 
the planetesimals succeeded to form at a wide range of distances from the star but, 
for some reasons, have not grown further to gas giants. Such a belt would undergo a 
long-term collisional erosion. This erosion is faster closer to the star, leading to 
a radial mass surface density profile $\Sigma(r)$ of the planetesimal disc rising outwards 
with a slope of about $7/3$ \citep[e.g.,][]{Kennedy2010,Wyatt2012}, although the exact 
value can be different under different assumptions for the initial radial profile of 
solids in the disc by the time of gas dispersal, critical fragmentation energy of 
planetesimals, and other parameters. 
We have checked the radial slope $\alpha_\Sigma$ of the mass surface density by evaluating
\begin{align}
 \alpha_\Sigma = \frac{\Delta \log \Sigma}{\Delta \log r} = 
 \frac{\log \Sigma(r_\text{o})-\log \Sigma(r_\text{i})}{\log r_\text{o}-\log r_\text{i}},
\end{align}
where $r_\text{i}$ and $r_\text{o}$ are the mean distance of the inner and outer planetesimal
belt, respectively. We found $\alpha_\Sigma\approx1.1$ for model~I 
and $\alpha_\Sigma\approx0.7$ for model~II, which are close to 1 and clearly confirm a 
solid density profile rising from the inside out. Also, the fact that $e_\text{max}$ in 
the outer component is lower than in the inner one is consistent with this collisional 
erosion scenario, because the biggest stirrers in the disc are smaller farther out from 
the stars, and their formation takes longer there \citep[e.g.,][]{Kenyon2008}.
We conclude that our models~I and II are consistent, at least qualitatively, with a long-term 
inside-out collisional erosion of an extended planetesimal disc.

\section{Prospects for observations}
\label{sec:prospects}

In the previous sections we presented several scenarios that can reproduce SED and 
radial profiles in the HIP~17439 system. We now test the capability of present-day 
observational facilities to reveal the actual debris disc structure.

\begin{figure}[htb!]
\resizebox{\hsize}{!}{\includegraphics{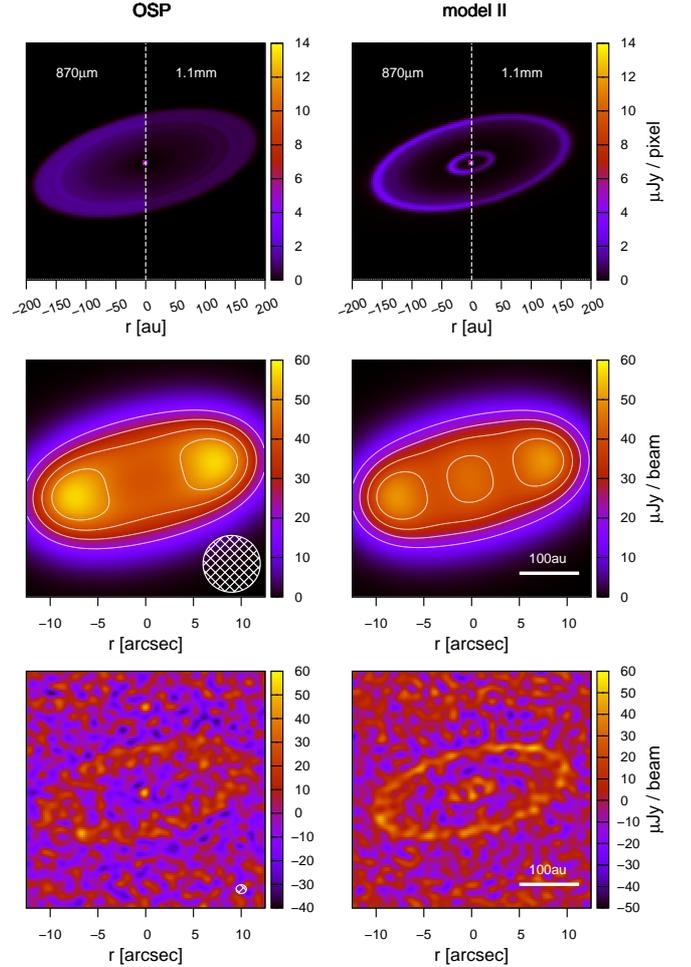}}          
\caption{Discs with one (OSP, Table~\ref{tab:input_onering}) and two 
         (model~II, Table~\ref{tab:input_two}) components 
         as seen by different telescopes.
         \emph{Top panels:} Unconvolved discs           
         at $870\microns$ and 1.1~mm (left- and
         right-hand side of each splitted image). Units are Jy/pixel, 
          where 1\,pixel=$0.2{''}\!\times 0.2{''}$. 
         \emph{Mid panels:} Simulated LMT/AzTEC images that were generated  
         by convolving the models  
         at 1.1~mm  with a Gaussian (FWHM=6''). 
         The white circle in the lower right of the left image illustrates the beam size.
         The contours in each image represent $2\sigma$, $3\sigma$, $4\sigma$, and
         $5\sigma$ levels. 
         \emph{Bottom panels:} Synthetic ALMA images in band 7 
         (central wavelength $870\microns$), using 
         antenna configuration \texttt{alma.out01} and a water vapour overburden 
         of 0.1~mm. 
         The beam size is $1.1{''}\!\times0.9{''}$. 
         With a on-source integration time of $15~\hour$ a 
         sensitivity of $\approx\!10\,\micro$Jy/beam was achieved.
}         
\label{fig:CASA_images}
\end{figure}

In a first test, we intended to predict how our best one- and two-belt model 
would be seen by a single-dish telescope. The 50-m Large Millimetre Telescope (LMT) 
atop Sierra Negra is the most powerful telescope with respect to high 
resolution imaging of HIP~17439's disc at millimetre wavelengths, although 
it is located at the northern hemisphere (latitude $\approx$~+19$^\circ$), 
and HIP~17439 (declination $\approx$~-38$^\circ$) is difficult to observe 
from this latitude. Inspired by the resolution of the LMT/AzTEC camera, 
we convolved images of OSP and model~II at $\lambda=1.1$~mm with a 
Gaussian profile with FWHM=6''. We assumed a signal-to-noise ratio (S/N) of 6 
and estimated the standard deviation $\sigma$ of the background fluctuation 
as the given S/N divided by the peak flux in the simulated observations. 
In Fig.~\ref{fig:CASA_images} the synthetic AzTEC image of model~II shows 
an inner emission above the $5\sigma$ level and gives a direct evidence for 
the presence of an inner planetesimal belt. However, the comparison between OSP 
and model~II shows that the difference in the inner disc emission is small, 
$\approx1\sigma$, and is not sufficient to clearly distinguish between them. 
To reach a significant detection of the inner belt, its emission should be differ 
by at least $3\sigma$ compared to the inner surface brightness dip of the one-belt 
scenario. This requires an S/N 3 times higher than assumed. Unfortunately, the 
LMT/AzTEC Photometry Mode Calculator\footnote{http://www.lmtgtm.org/?page\_id=832} 
predicts an observing time of 6~days (including all overheads) to reach S/N=6. 
Thus, a significant detection of an inner disc emission is impossible with LMT
in a reasonable time. 

In a second attempt, we simulated interferometric observations with the 
Atacama Large Millitmetre/submillimetre Array (ALMA). We used 
CASA~4.1.0\footnote{http://casa.nrao.edu/} (procedure \texttt{simalma}) 
to generate synthetic ALMA images. We assumed observations in band 7 
under good weather conditions (precipitable water vapor =~0.1~mm) and 
found that the inner ring in the two-belt model is barely detectable, 
although a long integration time of 15~h was applied 
(Fig.~\ref{fig:CASA_images}). Thus, it would be difficult, but not 
impossible, to distinguish between the one-belt and two-belt models 
considered. We stress that other two-belt configurations, which reproduce 
SED and radial profiles equally well as the ones found, may mitigate the 
challenge of this work. For instance, if the inner planetesimal belt 
lies slightly farther away from the star as in our model~II, the produced
dust will become colder and brighter at long wavelengths, favouring its  
detectability with ALMA. 
 
Furthermore, measurements of the radial surface brightness profiles in 
submillimetre/millimetre images will reveal the extension of the 
planetesimal belts. In this way, the hypothesis of a broad planetesimal 
disc as described in Sect.~\ref{sec:extended_belt} can be tested.

\section{Conclusions and discussion}
\label{sec:conclusion}
   
Our collisional modelling of the HIP~17439's debris disc does not
allow us to draw any strict conclusions as to the underlying architecture 
of the system. 
The data can be fitted well with a single disc or two separated discs, 
with the radial distribution of the planetesimals being narrow  
in both cases.
Furthermore, we argue that an extended planetesimal belt, approximated 
by a multicomponent disc, would reproduce the data as well.  
Below we discuss the astrophysical plausibility of all three models.

\begin{enumerate}
 \item {\em One parent belt}.   
Provided that a planetesimal belt is located 
at a distance $>\!120\au$ from the star, suggested by the one-component
model of \cite{Ertel2014}, dust material has to be transported efficiently
from the planetesimal location to the inner region of the system, otherwise the 
inner disc's surface brightness is too shallow and contradicts the 
observed radial profiles. Since Poynting-Robertson drag alone is not sufficient 
enough, we found a need for strong stellar winds. We identified  a wind 
strength of 15 times the solar value as the upper limit to prevent the radial 
profiles from being steepened too much if the inner edge of the planetesimal belt 
lies closer than $150\au$ to the star. 
The question remains whether HIP~17439 does possess winds of that strength. 
In our best run the planetesimals reside between 150 and 180~au, i.e.
they are much closer to the star than in the one-component scenario
of \cite{Ertel2014}. 
There, the outer ring edge was found at about 400~au, but could also be located
much further inside due to large uncertainties.
In general, planetesimals and dust production at very large stellar distances 
seem to be unlikely because of the increasingly long growth and stirring timescales 
of planetesimals \citep[e.g.,][]{Kenyon2008}. Hence, the moderate radial range of 
planetesimals in the one-belt model presented in this study
better fits in the debris formation theory.

\item {\em Two parent belts}. 
Without stellar winds of sufficient strength, the data are consistent with a 
two-ring disc, with a warm inner and a cold outer component. Many systems are 
believed to have a two-belt configuration since their SEDs can be well fitted 
by using two blackbody curves 
\citep[e.g.,][]{Matthews2010,Donaldson2013,Broekhoven-Fiene2013,Su2013}. 
Surveys of two component discs around stars of different spectral types highlight
that inner and outer component have median temperature values 
of $\approx\!190~\kelvin$ and $\approx\!60~\kelvin$, respectively 
\citep[e.g.,][]{Morales2011,Ballering2013}.
This yields a distance ratio of $r_\text{outer}/r_\text{inner}=(190/60)^2\approx10$, 
indicating a distinct gap between both components. One possible explanation of 
this result considers planets which have formed within the discs and split 
them up \citep[e.g.,][]{Ertel&Wolf&Rodmann2012}. In our best-fit two-belt 
model (model~II) the mean planetesimal distances for the inner and outer disc 
have a ratio of $\approx\!5$, i.e. close to what was found for many other systems.
By testing different maximum eccentricities $e_\text{max}$ of the inner and outer
planetesimal belt, we discovered that both components must have similar dynamical 
excitation of $e_\text{max}\approx0.04$ to be in good agreement with the radial 
profile data. One problem is that this level, especially in the inner belt, is 
lower than what is expected for the stirring by a planet \citep[e.g.,][]{Mustill2009}, 
suggesting that the gap between the two belts may be not populated by planets.
In that case, however, it would be difficult to explain what else, if not planets,
has cleared up the wide gap between the two belts.
Nevertheless, the problem can be mitigated by the assumption that possible 
planets in the gap are in nearly circular orbits and/or have low masses.

\item {\em One extended parent belt}. 
We discuss this possible disc architecture by using more than two belts 
adjacent to each other as a proxy. In principle, this could be the best 
model consistent with planet(esimal) formation theories. The optical thickness 
profile would be nearly constant over a wide radial range, starting to 
decrease beyond $\approx\!200\au$. The radial surface density profile of 
the underlying planetesimal disc is rising outwards, consistent with a 
long-term inside-out collisional erosion of such a disc. Therefore, from the 
point of view of collisional modelling we can also confirm the extended 
planetesimal belt hypothesis already proposed in \cite{Ertel2014}.
\end{enumerate}

More observations are required to discriminate between the competing 
scenarios discussed in this paper. LMT and ALMA are the most promising 
facilities to shed light on the actual structure of HIP~17439's debris 
disc. We simulated LMT and ALMA images of our best one- and two-belt models 
at millimetre/submillimetre wavelengths. These tests highlight that 
present-day telescopes are possibly capable to distinguish between a 
one- or a two-belt model but only with high observational effort.

\begin{acknowledgement}
We thank the reviewer for a speedy and constructive report that helped to 
improve the manuscript. CS, TL, and AVK acknowledge support by the 
\emph{Deut\-sche For\-schungs\-ge\-mein\-schaft} (DFG) through projects 
Kr~2164/10-1 and \mbox{Lo~1715/1-1}. SE thanks the French National Research 
Agency (ANR, contract ANR-2010 BLAN-0505-01, EXOZODI) and PNP-CNES for 
financial support. JPM and CE are partly supported by Spanish grant 
AYA 2011-26202.

\end{acknowledgement}


\end{document}